\documentclass[aps,superscriptaddress,showpacs,nofootinbib,twocolumn]{revtex4}

\usepackage{eucal} \usepackage{epsfig}

\begin{document}

\title{Magnetization of Planar Four-Fermion Systems}

\author{Heron Caldas} \email{hcaldas@ufsj.edu.br} \affiliation{Departamento de
  Ci\^{e}ncias Naturais, Universidade Federal de S\~{a}o Jo\~{a}o del
  Rei,\\ 36301-160, S\~{a}o Jo\~{a}o del Rei, MG, Brazil}

\author{Rudnei O. Ramos} \email{rudnei@uerj.br} \affiliation{Departamento de
  F\'{\i}sica Te\'orica, Universidade do Estado do Rio de Janeiro, 20550-013
  Rio de Janeiro, RJ, Brazil}

\begin{abstract}

We consider a planar system of fermions, at finite temperature and density,
under a static magnetic field parallel to the two-dimensional plane.  This
magnetic field generates a Zeeman effect and, then, a spin polarization  of
the system. The critical properties are studied from the Landau's free
energy. The possible observable consequences of the magnetization of planar
systems such as polymer films and graphene are discussed.

\bigskip\bigskip

Published: {\bf Phys. Rev. B 80, 115428 (2009)}

\end{abstract}


\maketitle

\section{Introduction}

{}Field theories in two spatial dimensions have long been recognized as
important for understanding several physical phenomena that can be well
approximated as planar ones, such as high temperature superconductors and the
fractional quantum Hall effect.  In the last years there has been an enormous
interest in  these two dimensional systems in condensed matter physics.  In
particular, a large number of important new phenomena discovered recently  in
condensed matter lies in this class. As examples we can cite the
metal-insulator transition (MIT)~\cite{MIT1} and graphene, an isolated single
atomic layer of carbon, in which electron transport is essentially governed by
Dirac's relativistic equation~\cite{Graphene}.  It is then feasible to treat
many of these planar (or quasi planar) systems in  condensed matter in terms
of quantum field theory models  for fermions in 2+1 dimensions (two space like
and one time like coordinates).  

Among the many field theory models useful for understanding a plethora of
phenomena in condensed matter systems, those that include a four-fermion
interaction have been extensively used in the context of studies of planar
systems. In particular, Nambu-Jona-Lasino type of models~\cite{rose} in 2+1
dimensions and, among  them, the Gross-Neveu (GN) model~\cite{GN}, have been
employed to study these  systems. {}For example, the GN model has been
considered as an appropriate model to study low energy excitations of high
temperature superconductors~\cite{liu}, while analogous models with generic
quartic  fermionic interactions have also been used to study quantum
properties of  graphene~\cite{son}.

In these planar systems, GN field theory models for fermionic interactions are
commonly used to study their symmetry properties, in particular chiral
symmetry breaking and restoration, at finite temperature and densities and
also in the presence of external  (magnetic) fields. In this paper we will be
interested in investigating the effects of how an ``asymmetrical doping'' can
affect the chiral symmetry, or the metal-insulator transition in a GN type of
model for a two-dimensional  system of fermions. An asymmetrical doping can be
defined as an imbalance  between the chemical potentials of the electrons with
the two possible spin  orientations (``up" $\equiv \uparrow$, and ``down"
$\equiv \downarrow$)  inserted in the system by a doping process. Since the
densities of the  $\uparrow$ and $\downarrow$ electrons are directly
proportional to their  chemical potentials, an asymmetrical doping is
equivalent to an asymmetry in the chemical potentials for spin up and spin
down electrons.  This chemical potential asymmetry can be produced by the
effect of an  external magnetic field parallel to the system's two-dimensional
plane.

A magnetic field applied parallel to the system's two-dimensional plane
couples only to the spin of the electrons, but not to the electrons orbital
motion. Therefore, Landau levels that would appear due to the coupling of a
(perpendicular to the plane) magnetic field to the electrons' orbital motion
do not appear here. Instead, the in-plane magnetic field generates an
intrinsic Zeeman effect which polarizes the system.  This is because at zero
magnetic field electrons with spin up and spin down have the same density, but
an in-plane magnetic field, due to the Zeeman effect, causes a difference
between the spin up and spin down densities. Recent studies in the context of
graphene~\cite{ezawa,folk} have suggested that the Zeeman effect can be very
important for the electronic properties of these systems. The intrinsic Zeeman
effect is thus useful to reveal the important role played by the spin degree
of freedom  of the electron and the polarization of the system~\cite{Dent}.

Here we will focus on the properties of a planar system modeled by a  GN
four-Fermi interacting model and study how a spin density asymmetry influences
on the chiral symmetry of the system, {\it i.e.} on the role of the  Zeeman
contribution to the system's magnetization.  The rest of this paper is
organized as follows.  In Sec. II, we review the GN model and the effect of
including an in-plane constant magnetic field and the Zeeman effect. In
Sec. III we evaluate the Landau's free energy (the effective potential) of the
model and determine how the chiral symmetry is affected by the spin density
asymmetry. In Sec. IV we determine the magnetic properties of the system.
{}Finally, in Sec. V we present our concluding remarks.

\section{The Model Action}

We start by considering a planar four-Fermi model describing interacting
fermions with a Lagrangian density given by

\begin{eqnarray}
{\cal L} [\bar{\psi},\psi] &=& \sum_{s=\uparrow,\downarrow} \bar{\psi}^s
\left( i  \hbar \partial_t - i \hbar v_F \vec{\gamma}. \vec{\nabla} \right)
\psi^s \nonumber \\ &+& \sum_{s=\uparrow,\downarrow}  \frac{\lambda}{2N}\hbar
v_F \left(\bar{\psi}^s\psi^s\right)^2\;,
\label{Lag4F}
\end{eqnarray}
where $\psi^s$ is a four-fermion field with $N$ flavors and $s$ is an internal
symmetry index  (spin) that determines the effective degeneracy of the
fermions.  In Eq.  (\ref{Lag4F}) a sum in the flavors is implicit. {}For
example, for materials such as polyacetylene or graphene, $N=2$. We will keep
$N$ general throughout this work for convenience.

In Eq. (\ref{Lag4F}), $\lambda$ is a  coupling term and $v_F$ is the {}Fermi
velocity. The gamma matrices are $4\times 4$ matrices and we follow the 
representation given e.g. in Ref. \cite{Park} for fermions in 2+1 dimensions. 
In this case, the model (\ref{Lag4F}) possesses a discrete chiral
symmetry, $\psi \to \gamma_5 \psi$, $\bar{\psi} \to - \bar{\psi} \gamma_5$,
with the $\gamma_5$ matrix defined as in \cite{Park}. This discrete chiral symmetry 
is the one considered in this paper, along with its breaking and restoration conditions.
Note that it is broken when a gap (or a nonvanishing  vacuum expectation value for
$\langle \bar{\psi} \psi \rangle$) is generated.

It is useful to rewrite the fermion interaction term in Eq. (\ref{Lag4F}) in
terms of a boson field $\Delta$, in which case ${\cal L} [\bar{\psi},\psi]$
becomes

\begin{eqnarray}
{\cal L}[\bar{\psi},\psi, \Delta] &=& \sum_{s=\uparrow,\downarrow}
\bar{\psi}^s  \left[ i  \hbar \partial_t - i \hbar v_F
  \vec{\gamma}. \vec{\nabla} -  \Delta \right] \psi^s \nonumber
\\ &-&\frac{N}{2 \hbar v_F \lambda}  \Delta^2\;.
\label{LagTLM}
\end{eqnarray}
$\Delta$ can be seen, for example, as representing the coupling of electrons
to a local value of the dimerization.  Equation (\ref{LagTLM}) is equivalent
to Eq. (\ref{Lag4F}) as can be easily verified by using the field equation for
$\Delta(x)$ and substituting  it back into the Lagrangian density,
re-obtaining the characteristic  four-Fermi interaction. 

Models of the type of Eq. (\ref{LagTLM}), or equivalently Eq. (\ref{Lag4F}),
are of the form of a Gross-Neveu model \cite{GN}, with applications found in
many areas of condensed matter physics. {}For example, in one space dimension,
Eq. (\ref{LagTLM}) is the equivalent of the Takayama--Lin-Liu--Maki (TLM)
model \cite{TLM}, the continuous version of the model proposed by Su,
Shrieffer, and Heeger (SSH) for polyacetylene~\cite{SSH,Review}, in the
adiabatic approximation, where lattice vibrations are neglected, and used to
study the metal-insulator transitions in general (see e.g. \cite{PRB} and
references therein).  

Let us now consider the application of a generic external magnetic field to
the system and its effects. It is convenient that to start by writing the
grand canonical partition function for the Lagrangian density model
(\ref{LagTLM}), 

\begin{equation}
Z=\int D \Delta \prod_{s} D \psi^{\dagger}D \psi \, \exp  \left\{ -
S_E[\bar{\psi},\psi,\Delta] \right\} ,
\label{partfun}
\end{equation}
where the Euclidean action $S_E[\bar{\psi},\psi]$, from the Lagrangian density
Eq. (\ref{LagTLM}), reads

\begin{eqnarray}
\lefteqn{ S_E[\bar{\psi},\psi,\Delta] =  \int_0^{\hbar \beta} d\tau \int d^2x~
  \left\{  \sum_{s=\uparrow,\downarrow} \bar{\psi}^s   \left[ \hbar
    \partial_\tau  \right. \right. } \nonumber \\ && \left. \left. + i \hbar
  v_F \gamma_1  \left(\partial_x +  i \frac{e}{c}\, A_{x}  \right)  +  i \hbar
  v_F \gamma_2  \left(\partial_y +  i \frac{e}{c}\, A_{y}  \right)
  \right. \right.  \nonumber \\ && \left. \left.  +  \Delta +  \gamma_0 \mu +
  \frac{\sigma_s}{2} \gamma_0 \, g \, \mu_B B_\parallel \right] \psi^s +
\frac{N}{2 \hbar v_F \lambda}  \Delta^2 \right\}\;,
\label{action}
\end{eqnarray}
where $\beta=1/(k_BT)$, $k_B$ is the Boltzmann constant, $\mu$ is the chemical
potential, $A_{x}$ and $A_{y}$ are the vector potential  components (
e.g. corresponding to a magnetic field perpendicular do the system's plane),
$B_\parallel$ is the magnetic field parallel to the system's plane and
$\sigma_s \gamma_0 \, g \, \mu_B B_\parallel/2$ is the corresponding Zeeman
energy term, with $\sigma_\uparrow =1$, $\sigma_\downarrow =-1$,  $g$ is the
spectroscopic Lande factor and $\mu_B$ is the Bohr magneton.  We must point
out that graphene samples have been recently studied in  strong magnetic
fields (up to 45 T)~\cite{Kim}. In these experiments  the measured effective
$g^*$ factors were found very close to the bare electron $g$ factor ($g=2$). 
In our work developed below, we will not assign a specific value for $g$
and we keep it also general (like $N$) for convenience.

By choosing a gauge where the three-dimensional vector potential is given,
for example, by
$\vec{A} = (0,B_\perp x,B_\parallel y)$, we see from Eq. (\ref{action}) that
$B_\perp$ couples to the orbital motion of the electrons and it will result in
the Landau levels for the system in this magnetic field. The parallel magnetic
field couples to the electrons' spin and produces the Zeeman energy term in
Eq. (\ref{action}). {}From the form of the Zeeman energy term in
Eq. (\ref{action}) we see that it can be added to the chemical potential, thus
defining an effective  chemical potential term in the action of the form,

\begin{eqnarray}
\sum_{s=\uparrow,\downarrow} \mu_s \bar{\psi}^s \gamma_0 \psi^s &=&
\sum_{s=\uparrow,\downarrow}  \left(\mu + \frac{\sigma_s}{2} \, g \, \mu_B
B_\parallel\right) \bar{\psi}^s \gamma_0 \psi^s \nonumber \\ &=&
\mu_{\uparrow} {\psi^\uparrow}^{\dagger}  \psi^{\uparrow} + \mu_{\downarrow}
   {\psi^\downarrow}^{\dagger}  \psi^{\downarrow}\;,
\label{asym}
\end{eqnarray}
where $\mu_\uparrow= \mu + \delta \mu$, and $\mu_\downarrow= \mu - \delta
\mu$, with $\delta \mu = g \, \mu_B B_\parallel/2$. The role of $\mu$ can be
interpreted as to account for the extra density of electrons that is supplied
to the system by the dopants, while $\delta \mu$ measures the amount of
asymmetry introduced and it is directly proportional to the in-plane applied
external magnetic field. As we explained in the introduction, in this work we
will be concerned with the effects of the Zeeman term, so from now on we take
$B_\perp=0$ and only consider a constant in-plane external magnetic field
$B_\parallel \equiv B_0$.

\section{The System's Effective Potential under External Effects}

In the applications with models of the GN type, we are mostly interested in
studying the effective potential, or Landau's free energy, for a constant
scalar field configuration $\Delta_c$, in which case chiral symmetry breaking
and dynamical fermion mass generation can be most conveniently studied.  Here,
we use the effective potential for $\Delta_c$ for studying how an asymmetry
between the up and down fermions' spins, generated by the constant in-plane
magnetic field,  will change the phase diagram of the model. Possible
phenomenological applications of these results for systems like for example
graphene and planar organic conductors, as polyacetylene, will then be
discussed.  

The effective potential is defined from the grand canonical partition function
(\ref{partfun}) by

\begin{eqnarray}
V_{\rm eff}=-\frac{1}{{\beta \cal V}} \ln Z \;,
\end{eqnarray}
where  ${\cal V}$ is the volume. Then, from the partition function
Eq. (\ref{partfun}) with (\ref{action}) and using Eq. (\ref{asym}), for a
constant background scalar field $\Delta_c$ and in the mean-field
approximation, which is the same as considering just the leading terms in a
$1/N$ expansion, or the large-$N$ approximation \cite{coleman,largeNreview},
we obtain that the effective potential $V_{\rm eff}$ is given by 

\begin{widetext}
\begin{eqnarray}
V_{\rm eff} (\Delta_c,T, \mu_{\uparrow}, \mu_{\downarrow})=  \frac{N}{2 \hbar
  v_F \lambda}  \Delta_c^2  - N k_B T \sum_{s=\uparrow,\downarrow} {\rm tr}
\sum_{n=-\infty}^{+\infty} \int \frac{d^2 p}{(2 \pi \hbar)^2} \ln
\left[\left(-i \omega_n +  \mu_{s} \right) -v_F\gamma_0
  \vec{\gamma}. \vec{p}-\gamma_0 \Delta_c \right] \;,
\label{Veff1}
\end{eqnarray}
\end{widetext}
where $\omega_n=(2n+1)\pi k_B T$ are the Matsubara frequencies for fermions.
Note that in the absence of asymmetrical doping, $\mu_{\uparrow}=
\mu_{\downarrow} = \mu$, or $\delta\mu=0$, Eq. (\ref{Veff1}) just reproduces
(in the natural units where $\hbar=k_B=v_F=1$) the effective potential of the
Gross-Neveu  model in 2+1 dimensions~\cite{rose}.

Performing the sum  over the Matsubara frequencies and taking the trace in
Eq. (\ref{Veff1}), we find 

\begin{eqnarray}
\lefteqn{V_{\rm eff}(\Delta_c,T, \mu_{\uparrow}, \mu_{\downarrow}) =
  \frac{N}{2 \hbar v_F \lambda}  \Delta_c^2 } \nonumber \\ &&- 2 N  k_B T
\int\frac{d^2p}{(2\pi \hbar)^2 }~ \left[ \beta E_p +   \frac{1}{2}\ln
  \left(1+e^{-\beta E_\uparrow^+}\right)  \right.\nonumber \\  && \left. +
  \frac{1}{2}\ln \left(1+e^{-\beta E_\uparrow^-}\right) +   \frac{1}{2}\ln
  \left(1+e^{-\beta E_\downarrow^+}\right)  \right. \nonumber \\ && \left. +
  \frac{1}{2}\ln\left(1+e^{-\beta E_\downarrow^-}\right) \right]\;,
\label{poteff}
\end{eqnarray}
where $E_{\uparrow,\downarrow}^{\pm} = E_p \pm \mu_{\uparrow,\downarrow}$ and
$E_p=\sqrt{v_F^2 p^2+\Delta_c^2}$.  In terms of the band structure,  we can
interpret the result seen in Eq. (\ref{poteff}) as like $\delta \mu$ has
lifted the degeneracy of the conduction and valence bands in the
matter part of $V_{\rm eff}$.

At zero temperature and chemical potential and in the absence of
the external magnetic field, $V_{\rm eff}$ becomes

\begin{eqnarray}
V_{\rm eff}(\Delta_c)=\frac{N}{ 2 \hbar v_F \lambda} \Delta_c^2 - 2 N \int
\frac{d^2p}{(2\pi \hbar)^2}~  \sqrt{v_F^2 p^2 +\Delta_c^2}\;.
\label{poteffTmu0}
\end{eqnarray}
By using a momentum cutoff $\Lambda$ to regulate the vacuum divergent term of
$V_{\rm eff}(\Delta_c)$ and by defining a renormalized coupling $\lambda_R$ as

\begin{equation}
\frac{1}{\lambda_R} = \frac{\hbar v_F}{N}  \frac{d^2 V_{\rm eff}(\Delta_c)}{d
  \Delta_c^2}\Bigr|_{\Delta_c=\Delta_0}\;,
\label{renor}
\end{equation}
where $\Delta_0$ is a renormalization point, that can be chosen, as usual,  as
given by the nontrivial minimum of the renormalized effective potential.

In terms of $\lambda_R$, the renormalized effective potential reads (after
subtracting an irrelevant field independent divergent vacuum term)

\begin{eqnarray}
V_{\rm eff,R}(\Delta_c)&=&\frac{N}{ 2 \hbar v_F \lambda_R} \Delta_c^2
\nonumber \\ &+&\frac{N}{\pi (\hbar v_F)^2 } \left(\frac{|\Delta_c|^3}{3}
-|\Delta_0|\Delta_c^2  \right)\;.
\label{poteffR}
\end{eqnarray}
The nontrivial minimum of $V_{eff,R}(\Delta_c)$ can now be easily found and it
is given by 

\begin{equation}
\Delta_0 = \frac{\hbar v_F \pi}{\lambda_R}\;.
\end{equation}
At $\Delta_c=\Delta_0$, the effective potential reads

\begin{eqnarray}
V_{\rm eff,R}(\Delta_c=\Delta_0)=  - \frac{N}{ (\hbar v_F)^2 }
\frac{\Delta_0^3}{6 \pi}\;,
\label{poteffDelta0}
\end{eqnarray}
which shows that $V_{\rm eff}(\Delta_c=\Delta_0) < V_{eff}(\Delta_c=0)=0$.
The non-trivial solution is then energetically preferable for the (undoped)
system, which then corresponds to a (dynamical) gap, {\it i.e.}, the  presence
of a chiral nonvanishing vacuum expectation value.

The effective potential at finite chemical potentials and in the zero
temperature limit, from Eq. (\ref{poteff}), is given by

\begin{eqnarray}
\lefteqn{V_{\rm eff,R}(\Delta_c,\mu_{\uparrow,\downarrow})= V_{\rm
    eff,R}(\Delta_c) } \nonumber \\ && +  N \Theta_1 \int_0^{p_{F}^\uparrow}
\frac{p~dp}{2\pi \hbar^2} (E_p - \mu_\uparrow)  \nonumber \\ &&+ N \Theta_2
\int_0^{p_{F}^\downarrow} \frac{p~dp}{2 \pi \hbar^2} \left( E_p -
|\mu_\downarrow| \right) \;,
\label{poteffmu}
\end{eqnarray}
where $V_{\rm eff,R}(\Delta_c)$ is given by Eq.~(\ref{poteffR}),
$\Theta_{1,2}=\Theta(\mu_{\uparrow,\downarrow}^2-\Delta_c^2)$ is a step
function, defined as $\Theta(x)=0$, for $x<0$, and $\Theta(x)=1$, for $x>0$,
and $p_{F}^{\uparrow,\downarrow}$ is the {}Fermi momentum of the
$\uparrow$($\downarrow$) fermions, 

\begin{equation}
p_{F}^{\uparrow,\downarrow}=\frac{1}{v_{F}}\sqrt{\mu_{\uparrow,\downarrow}^2
  -\Delta_c^2}\;.
\label{pf}
\end{equation}
By performing the momentum integration in Eq. (\ref{poteffmu}), we obtain

\begin{eqnarray}
\lefteqn{ V_{\rm eff,R}(\Delta_c,\mu_{\uparrow,\downarrow}) =  V_{\rm
    eff,R}(\Delta_c) } \nonumber \\  && + \frac{N\,\Theta_1}{4\pi (\hbar
  v_F)^2 }\left(-\frac{\mu_{\uparrow}^3}{3}-\frac{2}{3}|\Delta_c|^3
+\mu_{\uparrow}\Delta_c^2 \right) \nonumber \\  && + \frac{N\,\Theta_2}{4\pi
  (\hbar v_F)^2 }\left(-\frac{|\mu_{\downarrow}|^3}{3}-\frac{2}{3}|\Delta_c|^3
+|\mu_{\downarrow}|\Delta_c^2  \right) \;.
\label{poteff2}
\end{eqnarray}

Minimizing $V_{\rm eff,R}(\Delta_c,\mu_{\uparrow,\downarrow})$ with respect to
$\Delta_c$ yields again the trivial solution ($|\Delta_c|=0$) and

\begin{eqnarray}
\label{min2}
\lefteqn{ |\Delta_c|-\Delta_0 +  \frac{\Theta_1}{2}
  \left(\mu_\uparrow-|\Delta_c| \right) } \nonumber \\ &&+ \frac{\Theta_2}{2}
\left( |\mu_\downarrow| - |\Delta_c| \right)=0\;.
\label{min-mu}
\end{eqnarray}
Before continuing, lets us specialize to the symmetric limit, $\delta
\mu=0$. In this case Eq. (\ref{min-mu}) reads $|\Delta_c|-\Delta_0 +
\Theta(\mu^2-\Delta_c^2) \left(\mu-|\Delta_c| \right)=0$. To solve this
equation we need to know the critical chemical potential $\mu_c$ at which the
symmetry is restored.  This is found through the equation $V_{\rm
  eff,R}(\Delta_c=0,\mu_c)=V_{\rm eff,R}(\Delta_c=\Delta_0,\mu_c)$,  which
yields $\mu_c=\Delta_0$. Thus, the ground state of the symmetric  system is
characterized by 
\begin{eqnarray}
\label{gs1}
\Delta_c= \left\{
\begin{array}{cc}
\Delta_0, &{\rm for}~~\mu<\mu_c\;,\\ 0,&{\rm for}~~\mu \geq \mu_c\;.
\end{array}
\right.
\end{eqnarray}
At and beyond the doping level $\mu_c$, we note that the energy of the
fermions is linear, $E_p=v_F p$, characteristic of two dimensional gapless
systems~\cite{Schakel}.

\begin{figure}[htb]
  \vspace{0.5cm}
  \epsfig{figure=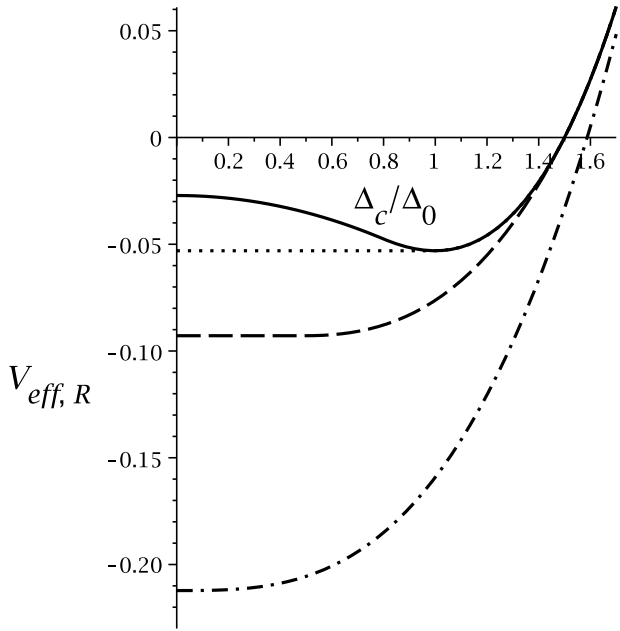,angle=0,width=8.5cm}
\caption[]{\label{Veff} The effective potential at zero temperature,
  Eq.~(\ref{poteff2}), in units of $N \Delta_0^3/(\hbar v_F)^2$.  The top
  curve (solid line)  is  for $\mu=0.8\mu_c$ and  $\delta \mu/\Delta_0=0$,
  with the minimum of $V_{\rm eff,R}$ at $\Delta_c=\Delta_0$. The dotted line
  is for $\mu=\mu_c$, and $\delta \mu/\Delta_0=0$, with the minimum of
  $V_{eff,R}$ at $\Delta_c=\Delta_0$ and at $\Delta_c=0$. The following curves
  are for $\delta \mu/\Delta_0=0.5$ (dashed line) and  $\delta
  \mu/\Delta_0=1.0$ (dash-dotted line).}
\end{figure}

Considering now the asymmetrical system case, we see that  by applying a
constant magnetic field $B_0$ parallel to the plane, the asymmetry $\delta
\mu$ is increased as $B_0$ increases. We study the asymmetry relative to
$\mu$, and look at the system's ground state  for a given asymmetry.   In
   {}Fig.~\ref{Veff} we show the effective potential $V_{\rm
     eff,R}(\Delta,\mu_{\uparrow,\downarrow})$ as a function of $\Delta_c$ for
   various asymmetries $0< \delta \mu/\Delta_0 <1$.  The first curve in
   {}Fig.~\ref{Veff} is the effective potential $V_{\rm eff,R}$ as a function
   of $\Delta_c/\Delta_0$  for $\mu=0.8\mu_c$ and $\delta \mu/\Delta_0=0$,
   with the minimum of $V_{\rm eff,R}$ at $\Delta_c=\Delta_0$. The second
   curve from top to bottom shows the symmetry restoration, where $\mu=\mu_c$,
   and $\delta \mu/\Delta_0=0$. The following curves are for
   $\delta \mu/\Delta_0=0.5$ and $\delta\mu/\Delta_0=1.0$. These results, in
   particular, show that there is  not a value for an asymmetry $\delta \mu$
for which a new minimum, where $\Delta_c(\delta \mu)\neq 0$, could appear. This is
   clearly seen from the  third and fourth curves in {}Fig.~\ref{Veff} (from
   top to bottom), which  show that the minimum of the effective potential is
   always at $\Delta_c=0$.  This result should be contrasted with the one
   obtained for the  one-dimensional space system case recently
   studied~\cite{NPB1}, where by increasing the asymmetry, a new minimum for
   the effective potential emerges at a critical chemical potential asymmetry
   $\delta \mu_c$. Thus, a non-vanishing and stable gap beyond $\delta \mu_c$
   would exist, till it disappears completely at $\delta \mu \gg \delta\mu_c$,
   through a second order phase transition.  The absence of this new minimum
   in our case will be explained below.
 
{}Finally, let us now consider the complete renormalized effective potential
at both finite chemical potential and temperature, obtained from
Eq. (\ref{poteff}) after performing the momentum integrals. It is given by

\begin{eqnarray}
V_{\rm eff}(\Delta_c,T, \mu_{\uparrow}, \mu_{\downarrow}) &=&
V_{\rm eff,R}(\Delta_c)  +  \frac{|\Delta_c|}{2 \pi \beta^{2}}  \left\{
      {\rm Li}_2[-e^{-\beta(\Delta_c-\mu_{\uparrow})}]  \right. \nonumber \\
      &+& \left.  {\rm
         Li}_2[-e^{-\beta(\Delta_c+\mu_{\uparrow})}]  \right\} \nonumber \\
       &+&  \frac{1}{2 \pi \beta^{3}}  \left\{
         {\rm Li}_3[-e^{-\beta(\Delta_c-\mu_{\uparrow})}]  \right. \nonumber \\
       &+& \left.   {\rm Li}_3[-e^{-\beta(\Delta_c+\mu_{\uparrow})}] \right\}
       + (\mu_{\uparrow} \to |\mu_{\downarrow}|)
\label{veffTmupm}
\end{eqnarray}
where  ${\rm Li}_\nu(z)$ is the polylogarithm function and it is defined (for
$\nu >0$) as \cite{Abramowitz}

\[
{\rm Li}_\nu(z)= \sum_{k=1}^{\infty} \frac{z^k}{k^\nu}\;.
\]

{}From Eq. (\ref{veffTmupm}) we can now verify how the Zeeman term, manifested
by the density asymmetry term $\delta \mu$, changes the usual (for $\delta
\mu=0$) chiral phase transition in the GN model when both finite chemical
potential and temperature are considered. We start by deriving  the gap
equation,

\begin{equation}
\frac{\partial}{\partial \Delta_c}V_{\rm eff}(\Delta_c,T, \mu_{\uparrow},
\mu_{\downarrow}) \Bigr|_{\Delta_c = \bar{\Delta}_c(T, \mu_{\uparrow},
  \mu_{\downarrow})} =0\;,
\label{gap}
\end{equation}
which gives

\begin{eqnarray}
\bar{\Delta}_c &=& \Delta_0 - \frac{1}{2\beta}\left\{ \ln \left[ 1+
  e^{-\beta(\bar{\Delta}_c+ \mu + \delta \mu)}\right] \right. \nonumber \\ &+&
\left. \ln \left[ 1+ e^{-\beta(\bar{\Delta}_c- \mu - \delta \mu)}\right] + \ln
\left[ 1+ e^{-\beta(\bar{\Delta}_c+ |\mu - \delta \mu|)}\right]
\right. \nonumber \\ &+& \left. \ln \left[ 1+ e^{-\beta(\bar{\Delta}_c- |\mu -
    \delta \mu|)}\right] \right\}\;.
\label{Deltacgap}
\end{eqnarray}
It can be easily checked that the $T=0$ limit of the above equation reproduces
the result (\ref{min2}).

\begin{figure}[htb]
  \vspace{0.5cm}
  \epsfig{figure=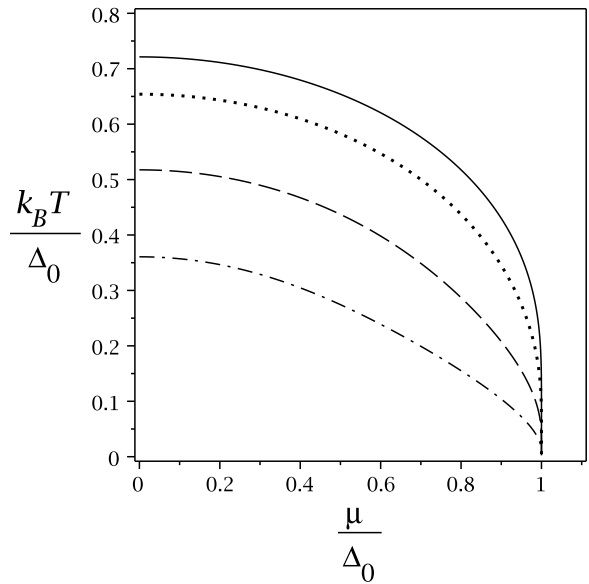,angle=0,width=8cm}
\caption[]{\label{critical} The critical curve $\bar{\Delta}_c(T,\mu,\delta
  \mu)=0$ for different values of asymmetry. The top curve (solid line) is for
  $\delta \mu=0$, while the other curves are for $\delta \mu = 0.5 \mu_c$
  (dotted line), $\delta \mu = 0.8 \mu_c$ (dashed line) and $\delta \mu = 0.95
  \mu_c$  (dash-dotted line), respectively.}
\end{figure}

The critical curve $\bar{\Delta}_c(T,\mu,\delta \mu)=0$, that is obtained from
Eq. (\ref{Deltacgap}), in the case $\delta \mu=0$, just reproduces the known
result~\cite{Park}, with a line for a  second order  phase transition in the
$\mu-T$ plane, starting at the critical point $(\mu=0,T=T_c)$, where $k_B T_c
= \Delta_0/(2 \ln 2)$,  and ending in a first order critical point at
$(\mu=\mu_c,T=0)$.  By increasing the asymmetry (the magnitude of the parallel
magnetic field $B_0$) the effect is to promote chiral symmetry restoration, as
can be explicitly seen in {}Fig. \ref{critical}.

We can now interpret the nonexistence here of new minima at $\mu=\mu_c$ in  the
presence of an  asymmetry, in contrast to the findings of Ref.~\cite{NPB1} for
the one space dimension case. This can be traced  to the nonexistence of a
critical line for first order chiral phase transition in the GN model in two
space dimensions in the mean-field  approximation.  The phase diagram of the
GN model in one space dimension, in the mean-field approximation, has a second
order critical transition line in the $\mu-T$ plane that meets a first order
transition line at a tricritical point. This is actually a typical phase
diagram seen in many other four-Fermi interacting models, including
Nambu-Jona-Lasino type of models in three space dimensions. However, this
typical phase diagram was absent in the two space dimensions GN model until
recently~\cite{gn3d}. In  Ref.~\cite{gn3d} it was shown that terms
contributing to the effective potential going beyond the mean-field
approximation would produce a first order critical line  merging with the
second order critical line at a tricritical point.  The effect of the
asymmetry close to the critical point $\mu_c$ seen in Ref.~\cite{NPB1} would then
probe the metastable region around the  first order critical line, manifested
by the formation of a new local minimum  in the presence of an asymmetry.  The
same should be seen here, if corrections beyond the mean-field approximation
would have been considered.  We do not see this formation of a new local
minimum close to the first order critical point here because  the mean-field
approximation in the two-space dimensional GN model misses this metastable
region.  We expect, though, that these corrections would produce  a very small
effect here.  This is because the metastable region seen in Ref. \cite{gn3d} was
considerable smaller than the one in the one-space dimensional GN model, so
its contributions to our derivations should, likewise, be small.

\section{Magnetic Properties}

As we have already seen, the imbalance in the chemical potentials of the
$\uparrow$ and $\downarrow$ electrons are induced by the application of a
static in-plane magnetic field in the system,  with a Zeeman energy $E_Z= \pm
g \mu_B B_0/2$ and, in the present case, $\delta \mu =|E_Z|=g \mu_B B_0/2$.
As a consequence, the number densities $n_{\uparrow}$ and $n_{\downarrow}$,
defined by
 
\begin{equation}
n_{\uparrow,\downarrow}= -\frac{\partial}{\partial \mu_{\uparrow,\downarrow}}
V_{\rm eff,R}(\bar{\Delta}_c,\mu_{\uparrow,\downarrow})\;,
\label{nud}
\end{equation}
are obviously imbalanced due to the asymmetry between $\mu_\uparrow$ and
$\mu_\downarrow$, and will depend on $\delta \mu$.  Likewise, the spin
polarization of the system as a result of the Zeeman effect produces a net
(Pauli) magnetization of the system, which is defined by~\cite{Kittel}

\begin{eqnarray}
M (T,\mu,\delta \mu )&=& \frac{g \mu_B}{2} (n_\uparrow - n_\downarrow)\;,
\label{mag}
\end{eqnarray}
and a magnetic (Pauli) susceptibility,

\begin{eqnarray}
\chi(T,\mu,\delta \mu ) &=& \frac{\partial}{\partial B_0} M(T,\mu,\delta \mu
)\;.
\label{magsusc}
\end{eqnarray}

Note that the polarization state of the system is determined by the asymmetry,
which depends on the intensity of the applied magnetic field.  As we change
the magnetic field, and then the polarization, we expect the magnetic
properties of the system. e.g. the magnetization and the susceptibility, to
change as well. As we are going to see, the change in behavior can be quite
remarkable, mostly when the susceptibility as a function of temperature is
concerned.  

\begin{figure}[htb]
  \vspace{0.5cm}
  \epsfig{figure=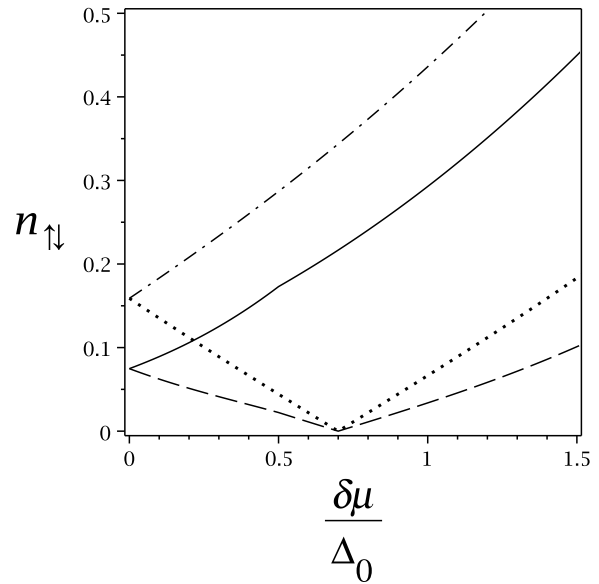,angle=0,width=8cm}
\caption[]{\label{nudfig} The number densities of spin up and down fermions as
  a function of the asymmetry, for $\mu=0.7 \Delta_0$ and for  temperatures
  $k_B T = 0.5 \Delta_0$ (solid and dashed lines, for  $n_\uparrow$ and
  $n_\downarrow$, respectively) and for  $k_B T= \Delta_0$ (dash-dotted and
  dotted lines, for $n_\uparrow$ and $n_\downarrow$, respectively).}
\end{figure}

Let us start by presenting some results for the number densities for fermions 
of spin up and spin down, as given by Eq. (\ref{nud}). We consider first the case of 
$\mu=0$ which represents the undoped system, {\it i.e.}, the insulating state at which 
$\Delta_c=\Delta_0$. Since $\mu=0$ we have effective chemical potentials given by 
$\mu_{\uparrow}= \delta \mu$ and $\mu_{\downarrow}= - \delta \mu$. 
{}From Eq.~(\ref{nud}) and the results of the previous section, one sees that if 
$\delta \mu < \Delta_0$, than $n_\uparrow = n_\downarrow = 0$, giving $M=\chi=0$. 
On the other hand, if $\delta \mu > \Delta_0$, there are non-vanishing densities, 
but they are always equal ($n_\uparrow = n_\downarrow $), resulting again in $M=\chi=0$. 
This means that the undoped (insulating) system, for which $\mu=0$, is never magnetized 
(polarized) at any temperature.
So, let us then consider the cases where $\mu \neq 0$. In {}Fig. \ref{nudfig}
we show how the spin up and down fermion densities  (in units of $N
\Delta_0^2/(\hbar v_F)^2$) change with the asymmetry ({\it i.e.}, when the
applied magnetic field increases), for a chemical potential chosen here, as
an example, as  $\mu
= 0.7 \mu_c \equiv 0.7 \Delta_0$ and for  temperatures  $k_B T = 0.5 \Delta_0$
(solid and dashed lines, for  $n_\uparrow$ and $n_\downarrow$, respectively)
and for  $k_B T= \Delta_0$ (dash-dotted and dotted lines, for  $n_\uparrow$
and $n_\downarrow$, respectively). Note that in both cases $n_\downarrow$
initially decreases with the increasing magnetic field, vanishing at the full
polarization point $\mu = \delta \mu$, as expected, and starts increasing
again beyond that point.  This is when the magnetic field becomes strong 
enough to be energetically favorable to spin down fermions of the filled 
valence band to be promoted to the conduction band. This occurs only at finite 
temperature, since at zero temperature $n_down$ is always zero at and beyond the 
critical magnetic field $B_c$ (or the full polarization point)~\cite{NPB1}.
Note that the spin up fermion density  always increase with the applied magnetic 
field (which is in the same direction of the spin up fermions).
The plots also indicate that the rate of spin down fermions turning into spin 
up ones (below the full polarization point) as the result of increasing the 
applied magnetic field is constant and is approximately the same rate of promotion 
(above the full polarization point) of spin down fermions of the valence band to the conduction band. Another result we can notice from {}Fig.~\ref{nudfig} is
that the  difference of the two densities, {\it i.e.} the Pauli magnetization
Eq. (\ref{mag}), is always positive with the increasing magnetic field.  This
is true for any other values of chemical potential and temperature.  

\begin{figure}[htb]
  \vspace{0.5cm}
  \epsfig{figure=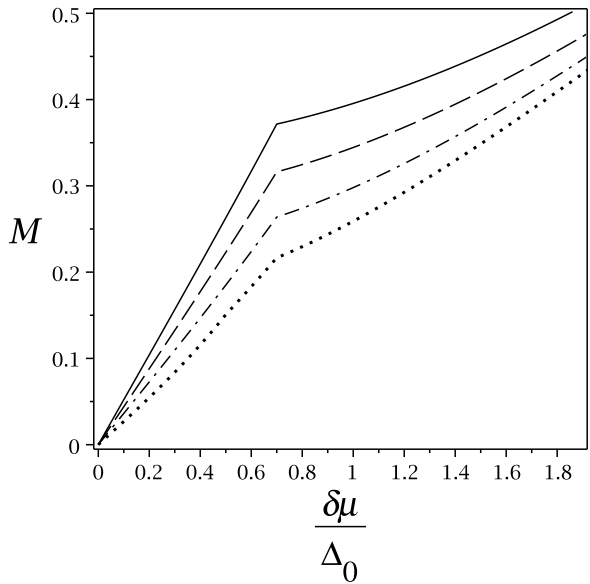,angle=0,width=8cm}
\caption[]{\label{magfig} The Pauli magnetization as a function of the
  asymmetry, for $\mu = 0.7 \Delta_0$ and for $k_B T =$ $0.5 \Delta_0$ (dotted
  line), $0.7 \Delta_0$ (dash-dotted line), $0.9 \Delta_0$ (dashed line) and
  $1.1 \Delta_0$ (solid line).}
\end{figure}

\begin{figure}[htb]
  \vspace{0.5cm}
  \epsfig{figure=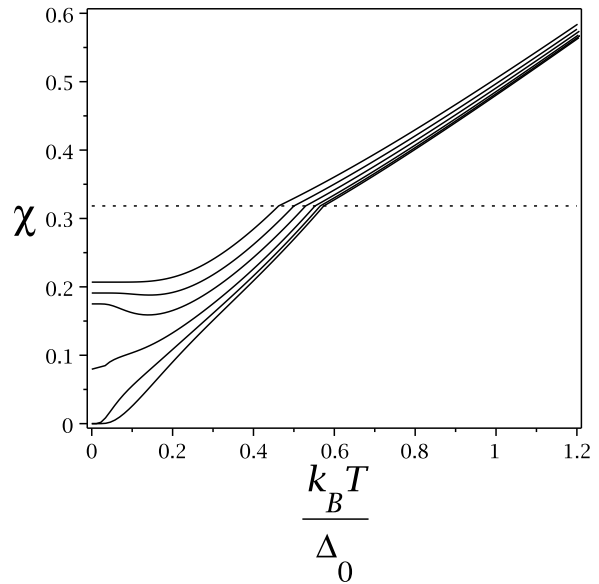,angle=0,width=8cm}
\caption[]{\label{chinotpolfig} The Pauli magnetic susceptibility as a
  function of the temperature for $\mu = 0.7 \Delta_0$,  for the values of the
  asymmetry $\delta \mu$ (from bottom to top) $0.1,\, 0.2,\, 0.3,\,  0.4,\,
  0.5,\, 0.6$ (in units of $\Delta_0$).  The dotted line indicates where the
  phase transition  happens.}
\end{figure}

\begin{figure}[htb]
  \vspace{0.5cm}
  \epsfig{figure=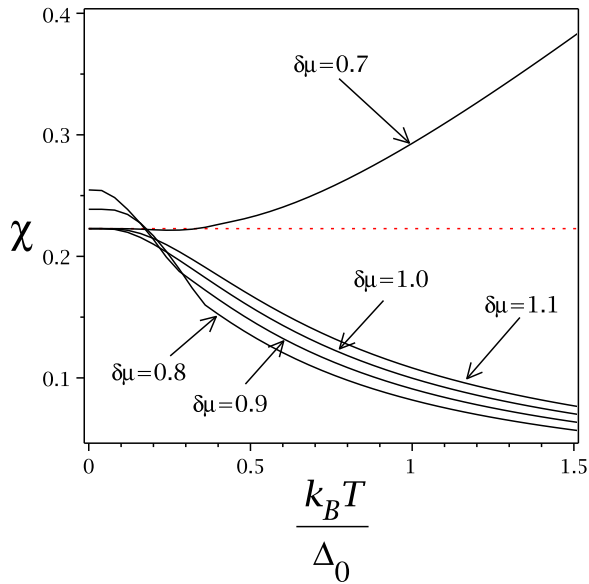,angle=0,width=8cm}
\caption[]{\label{chipolfig} The magnetic susceptibility as a function of the
  temperature for $\mu = 0.7 \Delta_0$, for the values of the asymmetry
  $\delta \mu=0.7,\, 0.8,\, 0.9,\, 1.0,\,  1.1$  (in units of $\Delta_0$).}
\end{figure}

Next, in {}Fig. \ref{magfig} we present the results for the Pauli
magnetization, expressed in units of $N (g \mu_B/2)\Delta_0^2/(\hbar v_F)^2$,
for different values of temperature (and again fixing the chemical potential
as $\mu = 0.7 \Delta_0$ for comparison purposes) when the asymmetry is
increased (or, equivalently, in terms of the Zeeman field $B_0$). Here we can
easily see a clear change in behavior of the magnetization as the full
polarization point is crossed. The most important observation we can notice
from the results seen in the {}Fig. \ref{magfig},  it is that below the full
polarization point, here given by  $\delta \mu = 0.7 \mu_c$ (which is the
value of $\mu$ that we have considered in the analysis), the larger is the
temperature, the larger is the rate of increase in the magnetization, while
has an opposite behavior above the full polarization point. This then will
reflect remarkably in the Pauli magnetic susceptibility results. As the
temperature is increased, the Pauli magnetic susceptibility should as well
increase for magnetic fields below of that which gives the full polarization.
Above the full polarization magnetic field, the magnetic susceptibility should
decrease with the increasing temperature.  This is confirmed by the results
presented in {}Figs. \ref{chinotpolfig} and \ref{chipolfig}, for the cases
below the full polarization point and at and above it, respectively.

{}Figures \ref{chinotpolfig} and \ref{chipolfig} show the   magnetic
susceptibility, expressed in units of  $N (g \mu_B/2)^2~\Delta_0/(\hbar
v_F)^2$, as a function of temperature, when the asymmetry is varied from
$\delta \mu = 0.1 \Delta_0$ up to $1.1 \Delta_0$, with fixed chemical
potential $\mu = 0.7 \Delta_0$. {}For all other values of chemical potential
we find similar behavior for $\chi(T,\mu,\delta \mu )$.

Analyzing the behavior of the magnetic susceptibility $\chi$ as a function of
the temperature, we find that, for those cases below the full polarization
point, e.g. {}Fig.~\ref{chinotpolfig},  in the chiral broken (insulating)
phase ($\bar{\Delta}_c \neq 0$), $\chi$ is nonlinear for $T<T_c$, where the
values of $T_c$ are indicated by the horizontal dotted line in
{}Fig.~\ref{chinotpolfig}. This horizontal critical line can be determined
explicitly and it is given by $N (g \mu_B)^2~\Delta_0/[4\pi (\hbar v_F)^2]$.
When this critical line is crossed and we go from the broken (insulating)
phase to the symmetric (metal) phase,  for $T \geq T_c$ (or $\bar{\Delta}_c
=0$), the magnetic susceptibility  becomes a linear function of the
temperature.  Note that the magnetic susceptibility, below the full
polarization point, is always an increasing function with the temperature, as
already anticipated by the behavior of the magnetization seen from
{}Fig. \ref{magfig}.

As we go above the full polarization point, the magnetic susceptibility
changes again its behavior with the temperature. We checked that at  high
temperatures, $k_B T \gg \Delta_0$, it goes exactly as predicted by the
Curie-Weiss law \cite{Kittel},  $\chi(\mu < \delta \mu, T \gg \Delta_0) \sim
1/T$. The turning point of behaviors for the magnetic susceptibility, as seen
in  {}Figs. \ref{chinotpolfig} and \ref{chipolfig}, is the full polarization
point $\mu = \delta \mu$.

A few analytical results for the magnetic susceptibility can be obtained
explicitly. At zero temperature, for any  $\mu+\delta \mu
<\mu_c$, we find that $\chi(T=0,\mu,\delta \mu) =0$, while for $\mu +\delta \mu \geq
\mu_c$ we find that $\chi(T=0,\mu,\delta \mu) \neq 0$. The two curves from bottom to top
seen in {}Fig. \ref{chinotpolfig} correspond to the former case, where
$\mu+\delta \mu <\mu_c$.  {}For the cases where $\mu +\delta \mu = \mu_c$,
with $\mu < \mu_c$ or $\delta \mu < \mu_c$, which in {}Fig. \ref{chinotpolfig}
corresponds to the third  curve from bottom to top, we find the analytical
result for the magnetic  susceptibility at $T=0$,

\begin{equation}
\chi(T=0,\mu+\delta \mu= \mu_c )\bigr|_{\mu < \mu_c,\delta \mu < \mu_c}  =
\frac{N (g \mu_B)^2}{16 \pi (\hbar v_F)^2 }~\Delta_0\;.
\label{chibrok}
\end{equation}

\noindent
At $\mu = \mu_c$, for any $\delta \mu$, or $\mu+\delta \mu \geq \mu_c$ such
that $\Delta_c=0$ (see {}Fig. \ref{Veff}),  {\it i.e.},  in the metal (chiral
restored) phase, we find that

\begin{equation}
\chi(T=0,\Delta_c=0)   = \frac{N (g \mu_B)^2}{4\pi (\hbar v_F)^2 }~\Delta_0\;.
\label{chires}
\end{equation}
This, in particular, corresponds to the values of parameters determining the
critical line seen in {}Fig. \ref{chinotpolfig}. 

The full polarization point can be reached for various dopings and in-plane
magnetic fields, { \it i.e.}, always that $\mu = \delta \mu$, in which case
$\mu_{\downarrow}=0$. In particular, at the full polarization point, for any
$\mu>0.5 \mu_c$, we find that

\begin{equation}
\chi(T=0,\mu>0.5 \mu_c, \mu=\delta \mu)  = \frac{N (g \mu_B)^2}{4\pi (\hbar
  v_F)^2 } \, \mu\;.
\label{chimu}
\end{equation}
We have also verified that this same result for the susceptibility  is also
obtained for other cases outside the full polarization point, e.g.  for $\mu
\geq \mu_c$, with any value of the asymmetry $\delta \mu$,  or for $\mu>0.5
\mu_c$ and $\delta \mu \geq \mu_c$.  Note that this result includes the case
leading to Eq.~(\ref{chires}),  for the special case of $\mu = \mu_c$.  The
curves starting at the dotted line in  {}Fig.~\ref{chipolfig} correspond
exactly to examples of these cases, with the dotted line obtained when $\mu =
0.7 \Delta_0$ is substituted in  Eq.~(\ref{chimu}).

The parameters determining the result given by Eq.~(\ref{chimu}),  together
with those determining Eq.~(\ref{chibrok}), show that there are two major
regions of parameters,  at zero temperature, that preclude the Pauli magnetic
susceptibility for attaining a value. We find that the Pauli magnetic
susceptibility jumps discontinuously from zero to the value given by
Eq.~(\ref{chibrok}),  for $\mu +\delta \mu = \mu_c$, with $\mu < \mu_c$ and
$\delta \mu < \mu_c$, and then it jumps again from the value given by
Eq.~(\ref{chibrok}) to the limiting lower value of susceptibility given by
Eq.~(\ref{chimu}), obtained for $\mu=0.5 \mu_c$.   {}For any other possible
value that the magnetic susceptibility can have, it will be larger than the
value obtained from Eq. (\ref{chimu}).  This behavior for the magnetic
susceptibility could in principle be tested experimentally, induced by either
a change in chemical potential  (e.g. by increasing the doping concentration
in planar systems) or  by an increase in the magnetic Zeeman field (thus
increasing the spin asymmetry). 

We also note that, as we have seen in the previous section, since for any
value of  $\mu < \mu_c$, the value of asymmetry  $\delta \mu = \mu_c$ will
lead to the phase transition. This value of asymmetry  corresponds to a
critical magnetic field for the chiral phase transition.  Since $\delta \mu_c
= \Delta_0 = g \mu_B B_{0,c}/2$, we find that this critical magnetic field is
given by 

\begin{equation}
B_{0,c} = \frac{2 \Delta_0}{g \mu_B}\;,
\label{cf}
\end{equation}
where we have used that $\mu_c = \Delta_0$. We may compare this result,  for
example, with that of MIT~\cite{MIT1,Pudalov}:

\begin{equation}
B_{pol} = \frac{2E_F}{g^* \mu_B}\;,
\label{cf2}
\end{equation}
where $E_F$ is the {}Fermi energy, $g^*$ is the effective Lande $g$ factor and
$m^*$ is the effective mass.

\section{Conclusions}

To summarize, we have investigated the mean-field phase diagram of planar
systems, that can be modeled with a four-Fermi type of model, upon asymmetric
doping. We have obtained the magnetization and magnetic susceptibility for
this system.  The analysis made in this work can be of relevance in studies of
many planar systems of interest in condensed matter, such as, for example,
organic conductors made of polymer films and graphene.  In particular,
regarding the zero temperature magnetic susceptibility of these systems, we
have predicted that it  can change abruptly from zero to the value given by
Eq.~(\ref{chibrok}), when increasing either the doping or the magnetic Zeeman
field, such that $\mu_\uparrow$ crosses a critical doping value given by
$\mu_c$. Then, it can jump again abruptly up to the point given by
Eq.~(\ref{chimu}), with  the minimum value attained for
$\mu = 0.5 \mu_c$.  This is a direct prediction of our results that could be
tested experimentally.

As far the dependence on temperature is concerned, we have obtained that
the magnetic susceptibility can have quite different behaviors depending
whether the system is below or above the full polarization point $\mu = \delta
\mu$. Below the full polarization point, $\mu < \delta \mu$,  while in the
chiral broken phase  the magnetic susceptibility increases nonlinearly with
the temperature, in the chiral symmetric phase it tends to a linear function
of the temperature. Unfortunately, it seems that  there is still so few or
none, to our knowledge, experimental results for the  magnetic susceptibility
of planar systems of the type that could be modeled  by the model studied
here, such as planar polymers or graphene, so to be able to compare with our
findings. Even so, despite the very different nature, we recall that there are
measurements of the behavior of the magnetic susceptibility with the 
temperature for metal alloys,
most notably titanium based~\cite{metalchi}, that displays exactly the same
behavior as we found here, increasing almost linearly with the temperature,
while the sudden change in the susceptibility at
the chiral phase transition point seems to be analogous to the  behavior seen
in the measured susceptibility of blue bronze (K$_{0.3}$MoO$_3$)
\cite{chandra} and indicative of a Peierls transition there.   The behavior of
the magnetic susceptibility changes again as we go above the full polarization
point. {}For $\mu < \delta \mu$ the magnetic susceptibility decreases when the
temperature increases. In particular, for high temperatures, $T \gg \Delta_0$,
it obeys the Curie-Weiss law. The applied magnetic field is then determinant
on the type of the behavior observed for the magnetic susceptibility. There is
a critical magnetic field, proportional to the chemical potential (the doping
concentration) that determines the two behaviors for the magnetic
susceptibility as a function of the temperature. We recall that there is a
similar behavior seen in pristine graphite~\cite{graphite}, where for low
magnetic fields the magnetic susceptibility is observed to increase with the
temperature, while for large fields it goes down with the temperature. The
major difference there with what we see here is that  graphite is diamagnetic,
while our results reflect the behavior of a paramagnetic material.  We hope
that in the future there will be measurements of the Pauli magnetic
susceptibility for planar systems so to be able to more closely compare with
the results we have obtained here.

Our results have also shown that the Zeeman effect in these planar systems
tends to weaken the chiral symmetry, thus, the insulating to metal
transition may happen at a smaller critical temperature in the presence of a
Zeeman field. This behavior, due to an increasing magnetic Zeeman field, is
exactly the opposite to what is observed when a perpendicular magnetic field
is applied to these systems (see e.g.~\cite{miransky,klimenko2}), in which
case the chiral symmetry breaking becomes stronger by the effect of the
magnetic field and, thus, the transition happens at a higher temperature in
the presence of a perpendicular magnetic field. These two opposite effects
caused by magnetic fields, when applied parallel or perpendicular to the
system's plane, can be an useful tool to regulate the insulating/metal
behaviors for these  planar systems, when the parallel and perpendicular
fields are applied simultaneously and independently, opening interesting 
possibilities for uses of these type of materials in practical applications 
as electronic devices. 
{}Further studies on the magnetic properties of these systems we hope
to make and to report on them  elsewhere in the future.

{}Finally, since in most realistic experiments the measured quantities of
interest are related to electric transport, it is expected also in those cases 
the Zeeman splitting to have important effects. However, a calculation of 
conductivity effects and transport properties cannot be addressed with the
methods we used here, based on the calculation of the effective potential 
(free energy), which are more suitable for the analysis of the phase structure of the
model. But based on experimental studies of an in-plane magnetic field, for example in 
graphene, the Zeeman splitting has been shown to be important in both the spin
transport and conductance fluctuations properties~\cite{folk}. It has also been shown
that the Zeeman splitting leads to the spectrum of the effective single-particle 
Hamiltonian exactly as required by the observed pattern of quantization of Hall 
conductivity~\cite{Kim}. These are interesting effects associated with the electronic 
transport properties of realistic systems, that we hope to present in the future and 
based on the model we have studied here.

\section{Acknowledgments}

We thank A. H. Castro Neto and I. Shovkovy for valuable discussions.
H. C. and R. O. R. are partially supported by CNPq.  H. C. also acknowledges
partial support from FAPEMIG.

\end{document}